\documentclass{appolb}
\usepackage{epsfig}
\usepackage{amsmath,amsfonts,amssymb}
\usepackage{t1enc}
\usepackage{cite}

\newcommand{\Dpp}{\Delta^{++}}
\newcommand{\Dmm}{\Delta^{--}}
\newcommand{\Dppmm}{\Delta^{\pm \pm}}
\newcommand{\Dp}{\Delta^+}

\newcommand{\Dmp}{\Delta^\mp}

\newcommand{\SM}{\Sigma_\text{M}}
\newcommand{\SD}{\Sigma_\text{D}}

\newcommand{\NM}{Z'_\lambda N_\text{M}}
\newcommand{\ND}{Z'_\lambda N_\text{D}}
\newcommand{\ptmiss}{p_T\!\!\!\!\!\!\!\!\not\,\,\,\,\,\,\,}

\begin{document}
\title{Trilepton signals: \\
the golden channel for seesaw searches at LHC
\thanks{Presented by F. del Aguila at ``NuFact09", 11th International 
Workshop on Neutrino Factories, Superbeams and Beta Beams, Chicago, 
Illinois (U.S.A.), July 20-25, 2009, and the XXXIII International Conference 
of Theoretical Physics ``Matter To The Deepest: Recent Developments in 
Theory of Fundamental Interactions", Ustro\'n, Poland, September 11-16, 
2009.}
}
\author{F. del Aguila, J.A. Aguilar-Saavedra and J. de Blas
\address{Departamento de F{\'\i}sica Te\'orica y del Cosmos and CAFPE, \\
Universidad de Granada, E-18071 Granada, Spain}
}
\maketitle
\begin{abstract}
The comparison of samples with different number of charged 
leptons shows that trilepton signals are the most significant ones 
for seesaw mediators. As previously pointed out, this is indeed the case for
scalar $\Delta$ (type II) and fermion $\Sigma$ (type III) triplets at LHC, 
which can be discovered in this channel for masses up to
$500-700$ GeV and an integrated luminosity 
of 30 fb$^{-1}$; whereas fermion singlets 
$N$ (type I) are marginally observable if there are no further 
new physics near the TeV scale. 
However, if there are new gauge interactions at this scale 
coupling to right-handed neutrinos, as in left-right models, 
heavy neutrinos are observable up to masses $\sim 2$ TeV for 
new gauge boson masses up to $\sim 4$ TeV, as we discuss 
in some detail. 
\end{abstract}

  
\section{Introduction}
Large hadron colliders can not directly test light neutrino masses 
because the energies they probe are of order of several hundreds of GeV, 
and then much larger than $m_\nu \sim 0.1$ eV.
However, they can be sensitive to them in definite models 
(see, for instance, \cite{Porod:2000hv}). 
In particular, they can produce the seesaw messengers generating 
the observed neutrino masses, if they have a mass near the electroweak 
scale $v \simeq 246$ GeV.

\begin{figure}[t]
\begin{center}
\begin{tabular}{ccc}
\epsfig{file=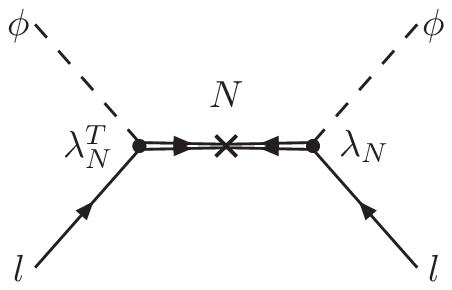,width=4.4cm,clip=} & 
\epsfig{file=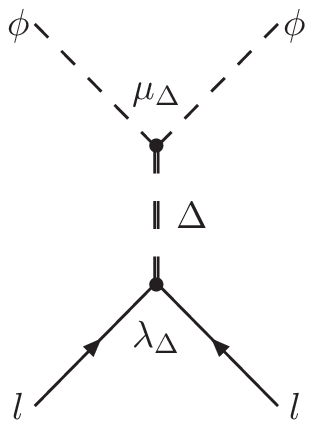,width=2.9cm,clip=} & 
\epsfig{file=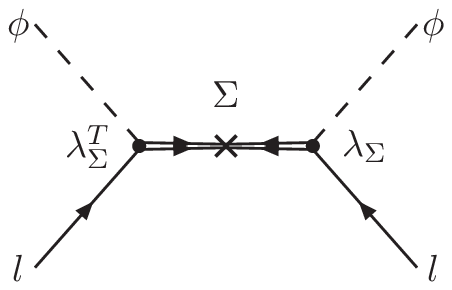,width=4.4cm,clip=} \\[1mm]
Type I & Type II & Type III \\[1mm]
$\frac{\alpha_5}{\Lambda}=\frac 12 \lambda_N^T m_N^{-1} \lambda_N$ &
$\frac{\alpha_5}{\Lambda}=-2 \frac{\mu_\Delta}{M_\Delta^2}\lambda_\Delta$ &
$\frac{\alpha_5}{\Lambda}=\frac 18 \lambda_\Sigma^T M_\Sigma^{-1} \lambda_\Sigma$ 
\end{tabular}
\end{center}
\caption{Tree level seesaw mechanisms.}
\label{seesaw mechanisms}
\end{figure}

The type I seesaw mechanism \cite{seesawI} 
invokes very heavy neutrino 
singlets $N$ slightly mixed with the Standard Model (SM) 
lepton doublets in order to explain the tiny neutrino masses 
observed in neutrino oscillation experiments. 
Their leading effects at low energy can be described by the 
lepton number violating (LNV) effective operator 
of dimension 5 resulting from the heavy neutrino integration 
\cite{Weinberg:1979sa} (see Ref. \cite{delAguila:2008ks} 
for notation and definitions)
\begin{equation} 
\frac{\alpha_5}{\Lambda}
{\cal O}_5= \frac{\alpha_5}{\Lambda}
\overline{L_L^{\ c}}\tilde \phi ^* \tilde \phi^\dagger L_L  
\rightarrow \frac{\alpha_5}{\Lambda}
\frac{v^2}{2}\overline{\nu^c} \nu \,.
\label{Op5}  
\end{equation} 
However, this operator can be also generated by the tree-level exchange 
of scalar $\Delta$ (type II seesaw) \cite{seesawII} and fermion $\Sigma$ 
(type III seesaw) \cite{seesawIII} triplets. 
In Fig. \ref{seesaw mechanisms} 
we gather the corresponding diagrams and 
the coefficients of the dimension 5 effective operator 
$\alpha_5/\Lambda$ 
resulting from the heavy particle integration.  
%
In order to reproduce the observed neutrino masses, 
$m_\nu \simeq (\alpha_5/\Lambda) v^2/2$,  
the effective coupling $\alpha_5$ must be 
quite small, of the order of $10^{-12}$ if 
the seesaw mediator has a mass $M = \Lambda$ near 
the electroweak scale (to be eventually 
at the LHC reach). This implies an 
effectively small lepton number violation at low energy: rather 
small $\lambda$ and/or $\mu$ (in the scalar case), 
and/or a large cancellation between different contributions. 
(See for further discussion and  
references~\cite{delAguila:2007ap,Integration}.)

If seesaw messengers are produced and detected at LHC, the mechanism 
for neutrino mass generation will be unveiled. 
The LHC discovery potential for seesaw mediators has been 
studied in detail for the minimal type I seesaw~\cite{delAguila:2007em}, 
as well as for extra charged~\cite{Ferrari:2000sp,Gninenko:2006br}
and neutral gauge interactions~\cite{delAguila:2007ua}; whereas 
simulations for type II seesaw have been performed in Ref.~\cite{delAguila:2008cj}, 
and for type III in Refs.~\cite{delAguila:2008cj,delAguila:2008hw,AguilarSaavedra:2009ik}. 
(Parton-level calculations can be found in Refs.\cite{partonI,partonIZ,partonII,partonIII}.) 
In general seesaw messenger production results in multilepton signals, 
with the significance of each final state depending not only on the signal 
cross section but also on its SM backgrounds. 
At this point, it is important to emphasize that there is no new physics process 
which is background free, even if it violates lepton number or flavour. 
For example, within the SM a final state with two charged leptons of the same-sign 
can be produced in association with two neutrinos balancing the total 
lepton number plus extra jets, as required by charge conservation. 
For instance,
\begin{equation}
u u \to W^+ W^+ d d \to \ell^+ \nu \ell^+ \nu d d \,,
\end{equation}
with $\ell=e,\mu$.
If the neutrinos have small transverse momenta their presence 
(via an observable missing energy $\ptmiss$) is unnoticed, and 
the process apparently violates lepton number. 
The same can be said about lepton flavour violating (LFV) 
final states, which can be mimicked by SM processes involving 
opposite-charge $W$ bosons. 
Apparently LNV backgrounds can be also produced if one charged 
lepton is missed by the detector, for example, in $WZ$ production 
with the lepton of different charge from $Z$ decay undetected.
A third, less trivial example is $t \bar t$ production with $\bar t$  
decaying semileptonically, 
\begin{equation}
q \bar q,gg \to t \bar t \to W^+ b \, W^- \bar b \to jj b \, 
\ell^- \bar{\nu} \bar b \,,
\end{equation}
and $b$ giving an isolated charged lepton, or viceversa.
There is some small probability that charged leptons from $b\to c \ell \nu$ decays 
have sizeable transverse momenta and small energy depositions in their vicinity, 
being not possible to effectively distinguish them in such a case 
from a charged lepton resulting from $W$ or $Z$ decay, 
except for their typically smaller transverse 
momenta. 
\footnote{Note that isolation criteria for electrons and muons must 
be relaxed at LHC experiments, allowing for a small amount of ``calorimeter noise'' 
in order to keep a good acceptance for leptons from $W$, $Z$ decays.}
Since the $t \bar t$ cross section is so large, this process is
a sizeable source of same-sign dileptons, 
being the dominant background in most cases. This makes compulsory 
to properly take it into account in the simulation.
A more detailed and enlightening discussion about how these backgrounds 
arise can be found in Refs.~\cite{delAguila:2007em,Sullivan:2008ki}.

As a general rule, it can be said that LNV signals, 
for instance same-sign dileptons $\ell^\pm \ell^\pm$, 
have much smaller backgrounds than lepton number conserving (LNC) signals 
{\em with equal number of charged leptons}, 
in this case $\ell^+ \ell^-$. 
(Signals which conserve lepton but violate flavour number, 
such as $\ell^+ {\ell'}^-$, have backgrounds of intermediate 
size.) 
But this does not apply when comparing signals and backgrounds with 
different charged lepton multiplicities, 
{\em e.g.}, $\ell^\pm \ell^\pm$, $\ell^\pm \ell^\pm \ell^\mp$ and
$\ell^+ \ell^+ \ell^- \ell^-$, as follows from simple arguments
and it is confirmed by detailed simulations.

In next section we show with some examples 
why trilepton signals are the best suited ones for discovery of type II 
and type III seesaw messengers, 
as well as of type I if heavy neutrinos couple to a new $Z'$ boson.  
This is due to their good sensitivity, the best one in most cases, 
for all seesaw models.
(This broad sensitivity in turn implies that the observation or not of other 
signals such as same-sign dileptons or four lepton final states is crucial 
to discriminate between models.) 
In Section~\ref{sec:3} we study multilepton signals from single heavy neutrino 
production in left-right (LR) models \cite{LR}, 
in the large parameter space region 
where heavy neutrinos predominantly decay into SM bosons $N \to lW / \nu Z / \nu H$. 
This process has not been previously studied in the literature, 
which focuses on the region where the three-body decay 
$N \to l W_R^* \to ljj$ and dilepton final states dominate
~\cite{Ferrari:2000sp,Gninenko:2006br}.

\section{Trilepton versus same-sign dilepton signals and backgrounds}

Dilepton and trilepton signals can appear in a variety of production processes 
involving seesaw messengers. In type I seesaw we can have single $N$ production
\begin{align}
& q \bar q' \to W^* \,/\, W' \to \ell^+ N \,,
\label{ec:lN}
\end{align}
with either a LNC decay $N \to \ell^- W^+$ or a LNV one $N \to \ell^+ W^-$. 
The subsequent $W$ boson decay results in only two leptons for $W \to q \bar q'$ 
or three for $W \to \ell \nu$. $N$ pair production is also possible if 
the heavy neutrinos couple to a new $Z'$ boson,
\begin{align}
& q \bar q \to Z' \to N N \,,
\label{ec:NN}
\end{align}
with $NN \to \ell^\pm W^\mp \, \ell^\mp W^\pm$ (LNC) or 
$NN \to \ell^\pm W^\mp \, \ell^\pm W^\mp$ (LNV). 
The fully hadronic decay $WW \to q \bar q' q \bar q'$ gives dilepton signals; 
whereas if one $W$ decays leptonically, three charged leptons are produced. 
In type II seesaw the processes
\begin{align}
& q \bar q \to Z^* \to \Delta^{++} \Delta^{--} \to l_1^+ l_2^+ l_3^- l_4^-
\,, \nonumber \\
& q \bar q' \to W^* \to \Delta^{++} \Delta^{-} \to l_1^+ l_2^+ l_3^- \nu \,,
\label{ec:DD}
\end{align}
with $l_i=e,\mu,\tau$, produce up to four charged leptons $\ell=e,\mu$. 
Finally, in type III seesaw we have, for example, 
\begin{align}
& q \bar q' \to W^* \to E^+ N \, ,
\label{ec:EN}
\end{align}
with $E^+ \to \ell^+ Z / \ell^+ H$  
and $N \to \ell^- W^+, \ell^+ W^-$ 
as in type I seesaw, and the subsequent decays of 
$Z,H \to q \bar q', \nu \bar \nu$ and $W$ into hadrons or leptons. 
In order to understand the relative significance and relevance of the 
different multilepton signals, several points have to be kept in mind:

{\em First.} Not all seesaw models involve heavy Majorana states and large lepton number 
violation. In particular, inverse type I, III seesaw models
~\cite{Inverse,delAguila:2007ap,delAguila:2008hw}
involve quasi-Dirac heavy neutrinos which in the processes in 
Eqs.~(\ref{ec:lN}), (\ref{ec:NN}) and (\ref{ec:EN}) do not produce final states with 
same-sign dileptons and no missing energy, but only opposite-sign ones. 
Still, trilepton signals do not require LNV neutrino decays $N \to \ell^+ W^-$, 
and are always present.

{\em Second.} In some cases, the branching ratio into three leptons 
is larger than into two same-sign leptons. For instance, in $Z' \to NN$ 
\begin{align}
& \text{Br}(\ell^\pm \ell^\pm \ell^\mp) \simeq \frac{1}{2} \times \frac{1}{2} \times \frac{2}{9}
\times \frac{6}{9} \times 2 \simeq  0.074 \,,   \nonumber \\
& \text{Br}(\ell^\pm \ell^\pm) \simeq \frac{1}{4} \times \frac{1}{4} \times 2  \times \frac{6}{9}
\times \frac{6}{9} \simeq  0.055 \,. 
\end{align}
Scalar triplet production and decay provides another example. 
In the light neutrino inverted mass hierarchy
the $\Dpp \to l_i^+ l_j^+$, $\Dp \to l_i \nu_j$ decays have 
approximate branching ratios
\begin{align}
& \text{Br}(\ell^+ \ell^+) \simeq \text{Br}(\ell^+ \nu_\ell) \simeq 0.65 \,, \nonumber \\
& \text{Br}(\ell^+ \tau^+) \simeq 2 \, \text{Br}(\ell^+ \nu_\tau)
\simeq 2 \, \text{Br}(\tau^+ \nu_\ell) \simeq 0.25 \,, \nonumber \\
& \text{Br}(\tau^+ \tau^+) \simeq \text{Br}(\tau^+ \nu_\tau) \simeq 0.1 \,.
\end{align}
(For studies of the dependence of these branching ratios on neutrino mixing parameters and the determination of neutrino data from collider observables see Refs.~\cite{recons}.)
Then, with a simple counting we obtain
\begin{align}
& \text{Br}(\Dpp \Dmm \to \ell^+ \ell^+ \ell^- \ell^-) \simeq 0.65 \times 0.65 \simeq 0.42
\,, \nonumber \\
& \text{Br}(\Dpp \Dmm \to \ell^\pm \ell^\pm \ell^\mp) \simeq 0.65 \times 0.25 \times 2 \simeq 0.32
\,, \nonumber \\
& \text{Br}(\Dpp \Dmm \to \ell^\pm \ell^\pm) \simeq 0.65 \times 0.1 \times 2 \simeq 0.13
\,, \nonumber \\
& \text{Br}(\Dppmm \Dmp \to \ell^\pm \ell^\pm \ell^\mp) \simeq 0.65 \times (0.65+0.12) \simeq 0.5
\,, \nonumber \\
& \text{Br}(\Dppmm \Dmp \to \ell^\pm \ell^\pm) \simeq 0.65 \times (0.125+0.1) \simeq 0.14 \,,
\end{align}
showing that trilepton final states dominate over four lepton and same-sign dilepton ones. 
For the normal hierarchy the trend is the same, 
also when secondary leptons from $\tau$ decays are included 
and simple event selection criteria imposed~\cite{delAguila:2008cj}.

{\em Third.} Even in processes where the branching ratio for 
$\ell^\pm \ell^\pm \ell^\mp$ is smaller than for $\ell^\pm \ell^\pm$ final states, 
as for instance in minimal type III seesaw (Eq.~(\ref{ec:EN})), 
the larger backgrounds in the latter case require more stringent cuts to reduce them 
making up for the difference in the signal cross sections. 
In order to illustrate these statements numerically,
we collect in Table~\ref{tab:nsnb-23} the number of same-sign dilepton and trilepton 
events evaluated with a fast detector simulation, and after typical selection cuts 
to enhance the signal significance for the processes in Eqs.~(\ref{ec:NN}) and (\ref{ec:EN}).
For comparison, we show both the Majorana and Dirac lepton triplet 
(labelled $\SM$ and $\SD$, respectively) signals, as well as the type I seesaw ones 
with an extra $Z'$ and a Majorana or Dirac heavy neutrino 
(labelled $\NM$ and $\ND$). We assume $m_{E,N} = 300$ GeV and 
$M_{Z'_\lambda} = 650$ GeV. 
It is apparent that for these cuts same-sign dilepton and trilepton backgrounds are quite similar, 
altough their relative size depends on the cuts applied.
In particular, we observe that the main background for same-sign 
dileptons (trileptons) comes from the semileptonic (dileptonic) channel 
in $t \bar t$ production, when a $b$ quark gives an isolated charged lepton.
As it has been already stressed in the introduction, 
the fact that a signal violates lepton number does not automatically 
guarantee the absence of SM backgrounds, 
nor imply that its background is much smaller than 
those for other LNC signals with more charged leptons. 
\begin{table}[htb]
\begin{center}
\begin{tabular}{l|cccl|cc}
        Signals       & $\ell^\pm \ell^\pm$ & $\ell^\pm \ell^\pm \ell^\mp$   & \quad &   Backgrounds           & $\ell^\pm \ell^\pm$ & $\ell^\pm \ell^\pm \ell^\mp$ \\[1mm]
\cline{1-3} \cline{5-7} 
& & & & & & \\ [-2mm]  
$E^+ E^-$ ($\SM$)     & 1.6   & 26.3  & & $t \bar t nj$         & 194   & 156 \\
$E^\pm N$ ($\SM$)     & 240.0 & 192.2 & & $tW$                  & 6     & 6   \\
$N N$ ($\NM$)         & 202.1 & 252.6 & & $W t \bar t nj$       & 12    & 47    \\
& & & &                                   $Z t \bar t nj$       & 3     & 20   \\
$E_i^+ E_i^-$ ($\SD$) & 4.2  & 80.9   & & $WW nj$               & 15   & 0     \\
$E_i^\pm N$ ($\SD$)   & 12.3 & 398.3  & & $WZnj$                & 24   & 38    \\
$N N$ ($\ND$)         & 8.1  & 481.9  & & $ZZnj$                & 4    & 5    \\
& & & &                                   $WWWnj$               & 7    & 12    \\
\end{tabular}
\end{center}
\caption{Number of events in the $\ell^\pm \ell^\pm$ and
$\ell^\pm \ell^\pm \ell^\mp$ final states for some signals and their main backgrounds 
and a luminosity of 30 fb$^{-1}$, from Ref.~\cite{AguilarSaavedra:2009ik}.}
\label{tab:nsnb-23}
\end{table}
We can also observe that, as indicated above,
in models with heavy Dirac neutrinos same-sign 
dileptons are practically absent, but trilepton signals 
are a factor of two larger than in the Majorana case. 
In next section we compare dilepton and trilepton final state 
production for the process mediated by the extra charged boson 
$W_R$ of a LR model in Eq.~(\ref{ec:lN}).

\section{Heavy neutrino production in left-right models at LHC}
\label{sec:3}

\begin{figure}[ht]
\begin{center}
\begin{tabular}{ccc}
\hspace{-1.2cm}
\epsfig{file=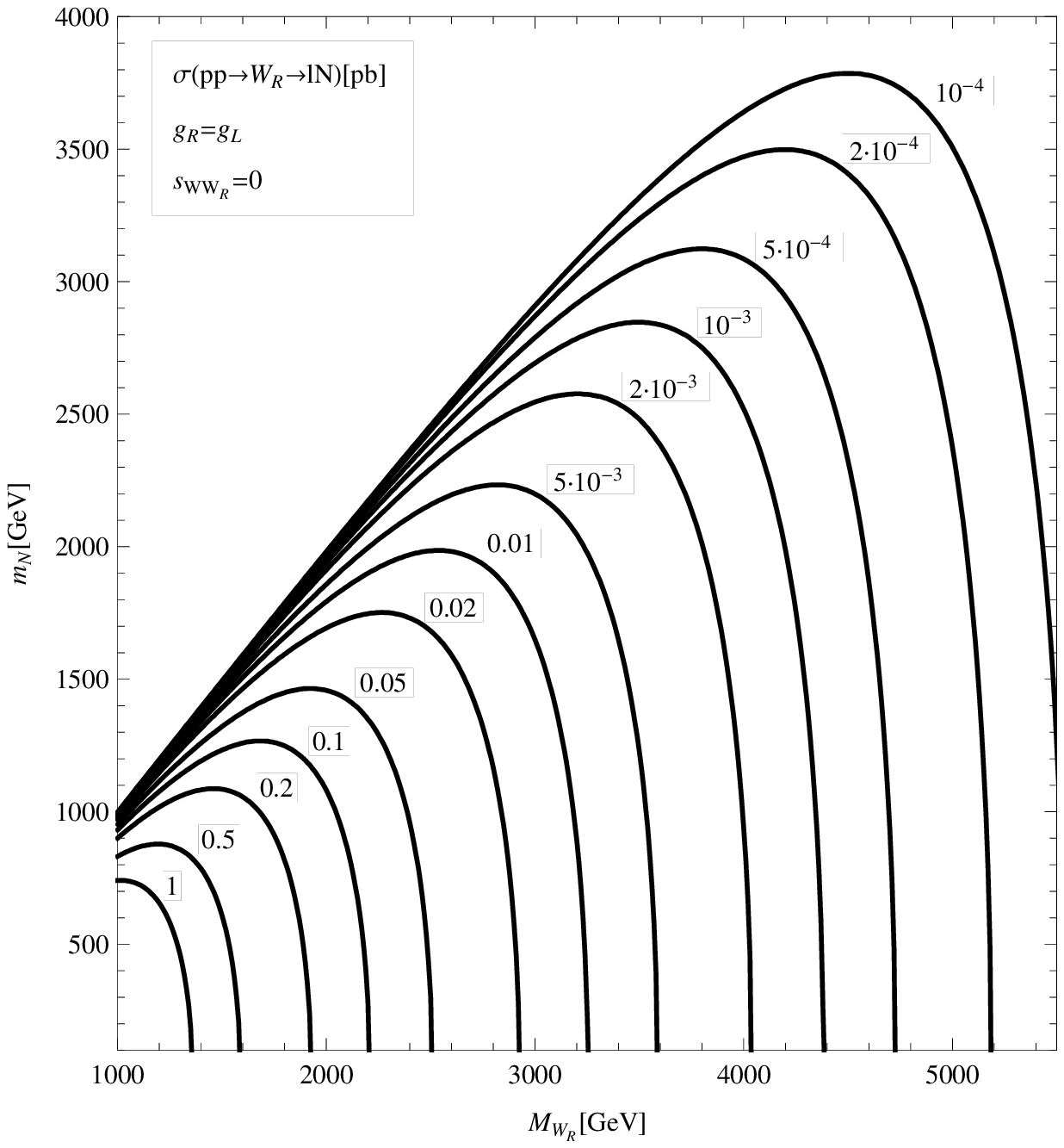,width=6.5cm,clip=} & \quad & 
\epsfig{file=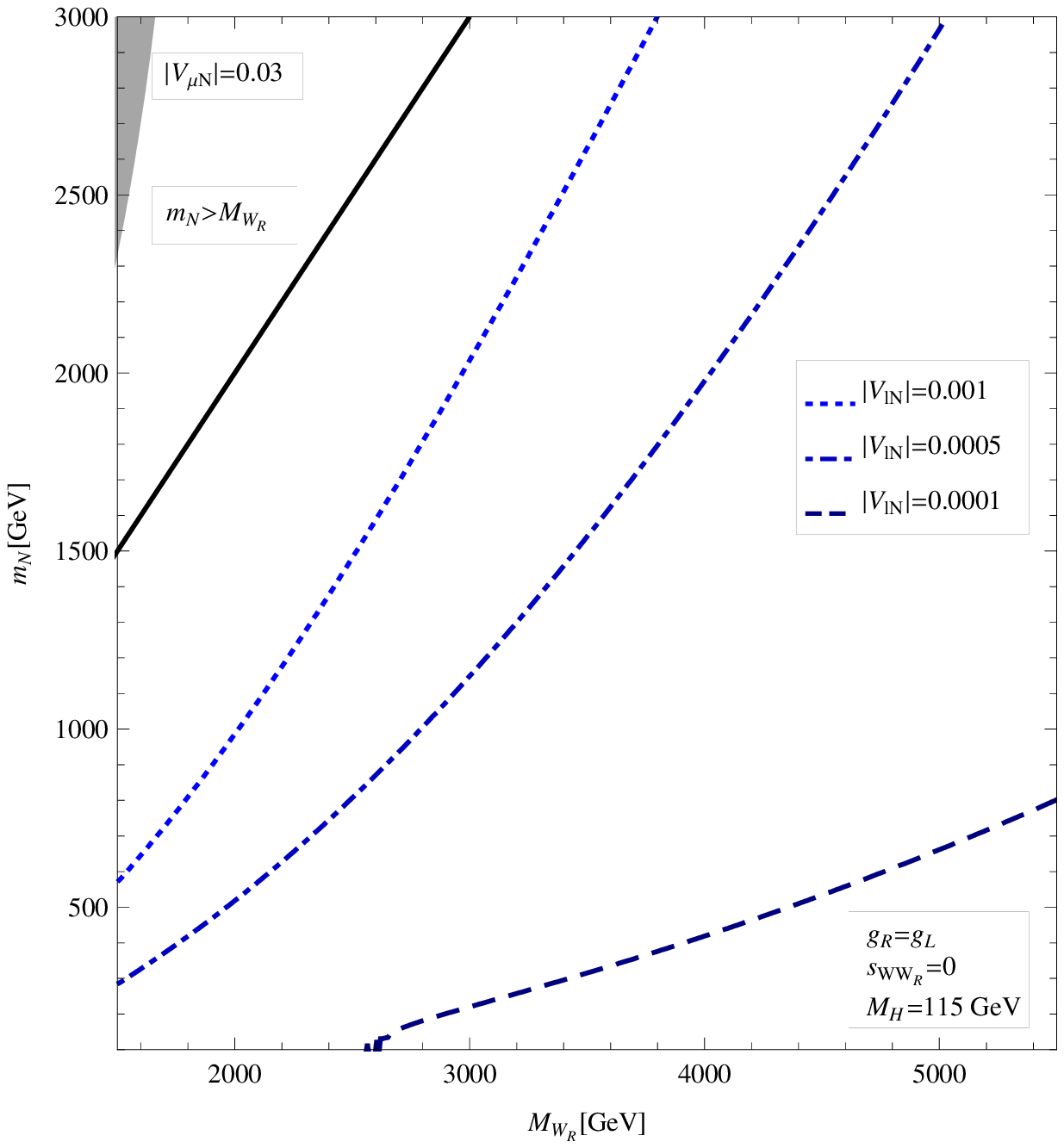,width=6.5cm,clip=}
\end{tabular}
\end{center}
\caption{Left: $pp\rightarrow W_R \rightarrow lN$ 
cross section at LHC. Right:
$N$ decay branching ratio to $W_R^*$ and SM bosons (see the text).}
\label{csbr} 
\end{figure}

Electroweak precision data constrain heavy neutrino singlets 
to mix little with SM leptons, $|V_{eN\ (\mu N)}| < 0.05\ (0.03)$ 
\cite{delAguila:2008pw}, 
\footnote{This new limit is derived including the 
CKM constraint, and recent data.} 
making them difficult to observe at LHC~\cite{delAguila:2007em}.
In models with extra $Z'$ bosons heavy neutrino pair production is possible, leading to dilepton and trilepton signals as shown in the previous section.
An alternative widely studied is the LR model, 
extending the SM gauge symmetry $SU(2)_L \times U(1)_Y$ to 
$SU(2)_L \times SU(2)_R \times U(1)_{B-L}$ and its 
matter content to include right-handed neutrinos $N$ 
\cite{LR}. 
Heavy neutrinos can then be produced by $W_R$ exchange 
with a relatively large cross section, as in Eq.~(\ref{ec:lN}), without any mixing suppresion.
The cross section, assuming $g_R = g_L$ and only one neutrino lighter than the $W_R$ boson, is 
plotted in Fig.~\ref{csbr} (left).
Available studies of the LHC reach for this process~\cite{Ferrari:2000sp,Gninenko:2006br} 
assume that $N$ only has a three-body decay $N \to l W_R^* \to ljj$.
However, this may not be the dominant mode for a 
relatively large range of LR model parameters. 
As a matter of fact, $N$ can mainly decay into $lW$, $\nu Z$, $\nu H$ 
if $V_{lN}$ or/and the $W-W_R$ mixing $s_{WW_R}$ are sizeable, of order $10^{-4}$ or larger.
The corresponding leptonic charged currents read  
\begin{eqnarray}                                                           
J_{W_R}^{-~\mu}& \simeq &-\frac{g_L}{\sqrt{2}}s_{WW_R}
\overline{\nu_L}\gamma^\mu l_L+\frac{g_R}{\sqrt{2}} 
\left(-V_{lN}\overline{\nu_L^{c}}\gamma^\mu l_R
+ \overline{N_R}\gamma^\mu l_R\right) \ ,  \cr
J_W^{-~\mu}& \simeq &\frac{g_L}{\sqrt{2}}\left(\overline{\nu_L}
\gamma^\mu l_L+V_{lN}\overline{N_R^{c}}\gamma^\mu l_L\right)
+\frac{g_R}{\sqrt{2}}s_{WW_R}\overline{N_R}\gamma^\mu l_R \ , 
\end{eqnarray}
where we only keep the leading terms in the small mixings and omit the flavour indices; 
and similarly for neutral currents. One can define the branching ratio
\begin{equation}
\text{Br}_R = 
\frac{\Gamma (N\rightarrow lW_R^*\rightarrow ljj)}
{\Gamma (N\rightarrow lW_R^*\rightarrow ljj)+
\Gamma (N\rightarrow lW, \nu Z, \nu H)} \ , 
\end{equation}
with 
\begin{equation}                                                         
\Gamma (N\rightarrow l W_R^*\rightarrow ljj) \simeq
N_c \frac{g_R^4}{1024\pi^3}\frac{m_N^5}{M_{W_R}^4} \ , 
\end{equation}
where we neglect quark masses and sum both 
lepton channels $\Gamma (N\rightarrow l^+ W_R^{- *}) = 
\Gamma (N\rightarrow l^- W_R^{+ *})$; whereas 
\begin{equation}
\Gamma (N\rightarrow l W) \simeq \frac{g_L^2\left|V_{lN}\right|^2+g_R^2
s_{WW_R}^2}{32\pi}\frac{m_N^3}{M_W^2}\left(1-\frac{M_W^2}{m_N^2}\right)^2\left(1+2\frac{M_W^2}{m_N^2}\right) , 
\end{equation}
and anagolously for $\nu Z, \nu H$ decays. In Fig. \ref{csbr} (right) we draw the curves for constant $\text{Br}_R = 1/2$ in the $M_{W_R}-m_N$ plane, which depend on the value of 
$|V_{lN}|^2+s^2_{WW_R}$. 
The curve corresponding to the present bound on $|V_{eN}|< 0.05$ (for $s_{WW_R}=0$) 
is in the $m_N > M_{W_R}$ upper-half out of the figure. 
For a given value of $|V_{lN}|^2+s^2_{WW_R}$, 
the region on the left of the curve corresponds to $\text{Br}_R > 1/2$, where $W_R^*$ 
decays start to dominate.

\begin{figure}[ht]
\begin{center}
\begin{tabular}{ccc}
\hspace{-1.2cm}
\epsfig{file=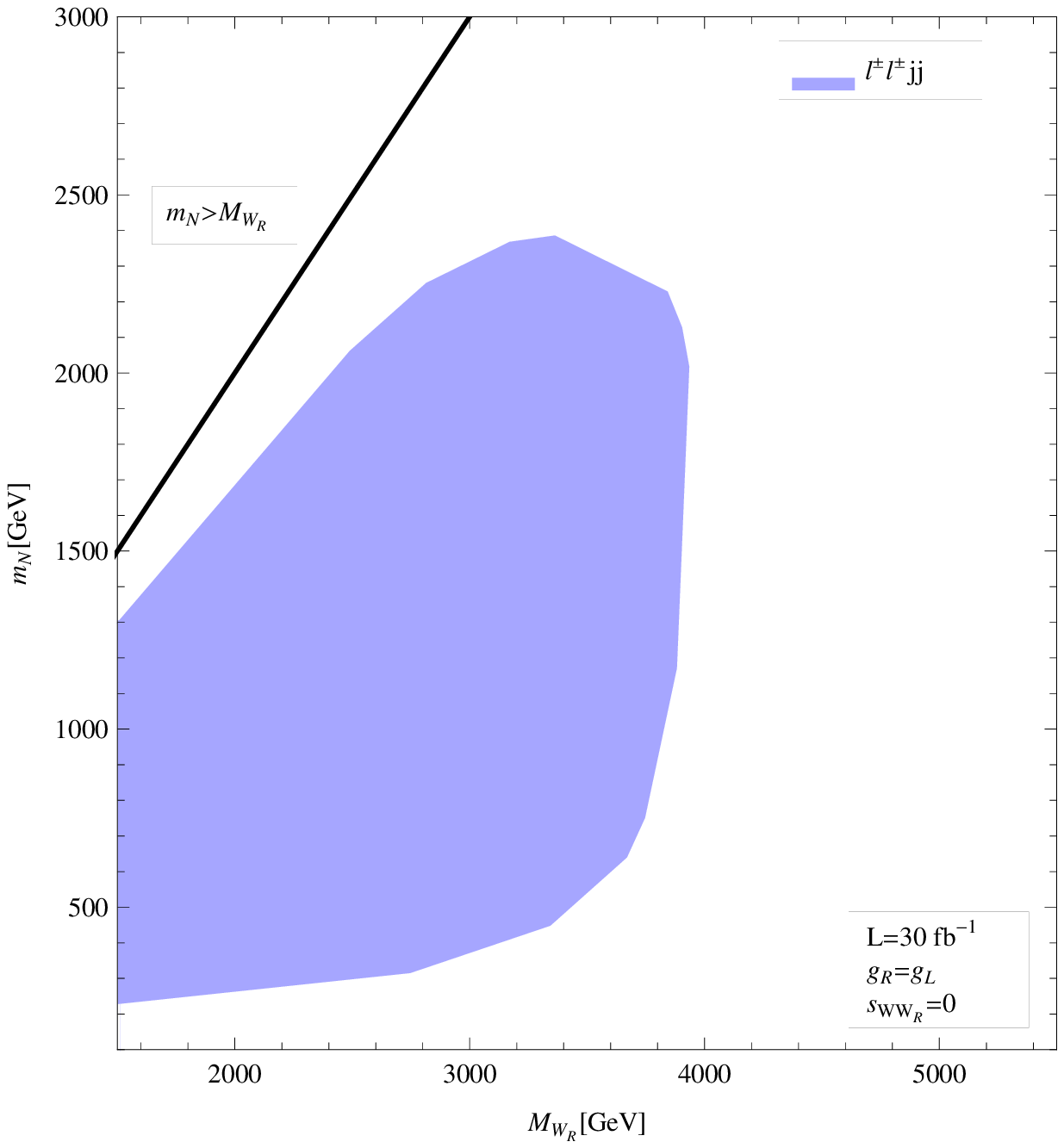,width=6.5cm,clip=} & \quad & 
\epsfig{file=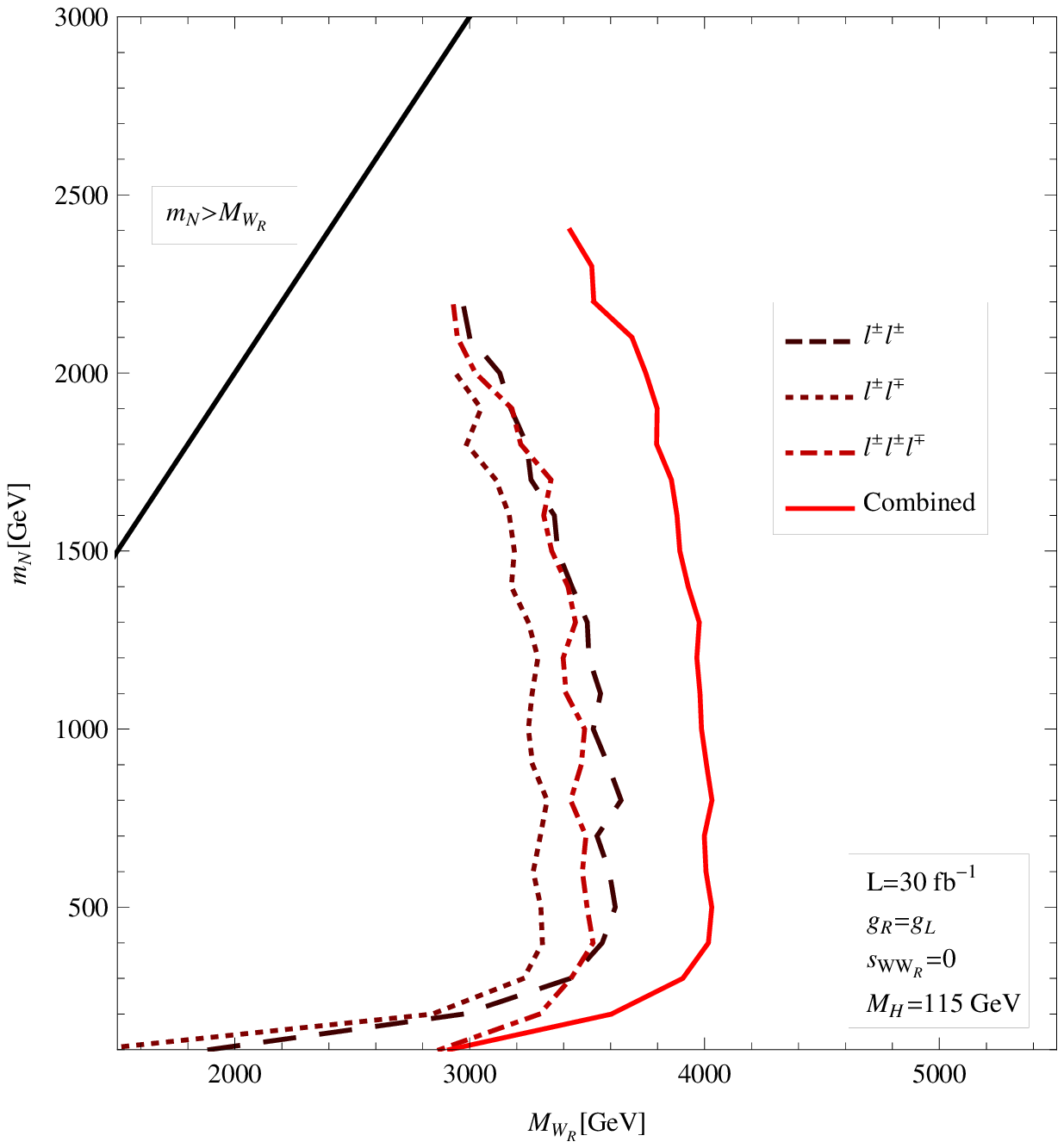,width=6.5cm,clip=}
\end{tabular}
\end{center}
\caption{Left: Discovery limits for 30 fb$^{-1}$ as a function of $M_{W_R}$ and $m_N$, assuming that 
$N$ only decays into $l^\pm W_R^* \rightarrow ljj$, from Ref.~\cite{Gninenko:2006br}.
Right: The same, assuming that $N$ decays into SM bosons.}
\label{DiscLimit} 
\end{figure}

The scenario with $\text{Br}_R \sim 1$ has been widely studied, and we present in Fig.~\ref{DiscLimit} (left) the limits obtained in Ref.~\cite{Gninenko:2006br}, assuming a 
100\ \% branching ratio into $N \to ljj$. For the scenario with $\text{Br}_R \sim 0$ we have performed a new simulation extending the generator {\tt Triada}~\cite{delAguila:2008cj} with this process and using {\tt Alpgen}~\cite{Mangano:2002ea} to generate the SM backgrounds. The parton shower Monte Carlo {\tt Pythia} 6.4\cite{Sjostrand:2006za} is used to add initial and final state radiation and pile-up, and perform hadronisation. The fast detector simulation {\tt AcerDET}~\cite{RichterWas:2002ch} is used to simulate a generic LHC detector. Analyses have been performed for different neutrino masses in steps of 
100 GeV, using $W_R$ masses of 2.5 or 3 TeV, close to the limits obtained. The 
selection criteria (with small modifications at some points) are:
\begin{itemize}
\item[]$\ell^\pm \ell^\pm \ell^\mp$: three leptons, the same-sign pair with 
$p_T > 30$ GeV and the third one with $p_T > 10$ GeV; invariant mass of 
opposite-sign pairs $|m_{\ell^+ \ell^-} - M_Z| > 10$ GeV; total invariant mass 
(adding the missing momentum) $m_\text{tot} > 1.5$ TeV.
\item[] $\ell^\pm \ell^\pm$: two same-sign leptons with $p_T > 30$ GeV; two jets 
with $p_T > 20$ GeV;
missing energy $\ptmiss < 50$ GeV; leading lepton with $p_T > 400$ GeV; 
$m_\text{tot} > 1.5$ TeV.
\item[] $\ell^+ \ell^-$: two opposite-sign leptons with $p_T > 30$ GeV and 
$m_{\ell^+ \ell^-} > 500$ GeV; two jets with $p_T > 20$ GeV; leading lepton 
with $p_T > 750$ GeV and leading jet with $p_T > 200$ GeV; $\ptmiss < 50$ GeV; 
$m_\text{tot} > 2.5$ TeV.
\end{itemize}
The limits for the dilepton and trilepton final state are presented in 
Fig.~\ref{DiscLimit} (right), as well as their combination. 
We observe that same-sign dilepton and trilepton signals have very similar significances despite the larger branching ratio into the former. This is because 
dilepton backgrounds are also larger and 
the signal efficiency for $m_N \ll M_{W_R}$ with a highly boosted 
$N$ is smaller in the dilepton channels, in which the charged lepton from $N$ decay is often located inside the jets from $W$ hadronic decay. The combination of all final states 
gives a discovery reach similar to the scenario
with $W_R^*$ decaying dominantly into two jets. 

\vspace{0.5cm}
We thank discussions with M. P\'erez-Victoria. 
This work has been supported by MEC project FPA2006-05294,
Junta de Andaluc{\'\i}a projects FQM 101 and FQM 03048, 
and by the European Community's Marie-Curie Research Training
Network under contract MRTN-CT-2006-035505 ``Tools and Precision
Calculations for Physics Discoveries at Colliders''.
J.A.A.-S. acknowledges support by a MEC Ram\'on y Cajal contract.


\begin{thebibliography}{99}

\bibitem{Porod:2000hv}
  W.~Porod, M.~Hirsch, J.~Romao and J.~W.~F.~Valle,
  {\it Phys.\ Rev.} {\bf D63}, 115004 (2001) 
  {\tt  [hep-ph/0011248].}

\bibitem{seesawI} M. Gell-Mann, P. Ramond and R. Slansky,
in {\it Supergravity}, ed. by P. Van Nieuwenhuizen and
D.Z. Freedman (North Holland, Amsterdam, 1979), p.315;
S.L. Glashow, in
{\it Quarks and Leptons}, Cargese, 1979,
ed. by M. Levy et al. (Plenum, New-York, 1980);
T. Yanagida, in Proc. of the
{\it Workshop on the Unified Theory and Baryon Number in the Universe},
ed. by O. Sawada and A. Sugamoto (KEK report 79-18,
Tsukuba, Japan, 1979), p.95;
R.N. Mohapatra and G. Senjanovic,
{\it Phys.\ Rev.\ Lett.} {\bf 44}, 912 (1980).

\bibitem{Weinberg:1979sa}
  S.~Weinberg,
  {\it Phys.\ Rev.\ Lett.} {\bf 43}, 1566 (1979).

\bibitem{delAguila:2008ks}
  F.~del Aguila, J.~A.~Aguilar-Saavedra, J.~de Blas and M.~Perez-Victoria,
  {\tt arXiv:0806.1023 [hep-ph]}


\bibitem{seesawII}
  M.~Magg and C.~Wetterich,
  {\it Phys.\ Lett.} {\bf B94}, 61 (1980);
  T.~P.~Cheng and L.~F.~Li,
  {\it Phys.\ Rev.} {\bf D22}, 2860 (1980);
  G.~B.~Gelmini and M.~Roncadelli,
  {\it Phys.\ Lett.} {\bf B99}, 411 (1981);
  G.~Lazarides, Q.~Shafi and C.~Wetterich,
  {\it Nucl.\ Phys.} {\bf B181}, 287 (1981);
  R.~N.~Mohapatra and G.~Senjanovic,
  {\it Phys.\ Rev.} {\bf D23}, 165 (1981).


\bibitem{seesawIII}
  R.~Foot, H.~Lew, X.~G.~He and G.~C.~Joshi,
  {\it Z.\ Phys.} {\bf C44}, 441 (1989);
  E.~Ma,
  {\it Phys.\ Rev.\ Lett.} {\bf 81}, 1171 (1998) 
  {\tt [hep-ph/9805219]}.


\bibitem{delAguila:2007ap}
  F.~del Aguila, J.~A.~Aguilar-Saavedra, J.~de Blas and M.~Zralek,
  {\it Acta Phys.\ Polon.} {\bf B38}, 3339 (2007) 
  {\tt[arXiv:0710.2923 [hep-ph]]};
  X.~G.~He, S.~Oh, J.~Tandean and C.~C.~Wen,
  {\tt arXiv:0907.1607 [hep-ph]}.

\bibitem{Integration}
  S.~Antusch, C.~Biggio, E.~Fernandez-Martinez, M.~B.~Gavela and J.~Lopez-Pavon,
  {\it J.\ High\ Energy\ Phys.} {\bf 0610}, 084 (2006) 
  {\tt [hep-ph/0607020]};
  A.~Abada, C.~Biggio, F.~Bonnet, M.~B.~Gavela and T.~Hambye,
  {\it J.\ High\ Energy\ Phys.} {\bf 0712}, 061 (2007) 
  {\tt [arXiv:0707.4058 [hep-ph]]}.


\bibitem{delAguila:2007em}
  F.~del Aguila, J.~A.~Aguilar-Saavedra and R.~Pittau,
 {\it J.\ High\ Energy\ Phys.} {\bf 0710}, 047 (2007) 
 {\tt [hep-ph/0703261]}.

\bibitem{Ferrari:2000sp}
  A.~Ferrari et al.,
  {\it Phys.\ Rev.} {\bf D62}, 013001 (2000). 

\bibitem{Gninenko:2006br}
  S.~N.~Gninenko, M.~M.~Kirsanov, N.~V.~Krasnikov and V.~A.~Matveev,
  {\it Phys.\ Atom.\ Nucl.} {\bf 70}, 441 (2007). 

\bibitem{delAguila:2007ua}
  F.~del Aguila and J.~A.~Aguilar-Saavedra,
  {\it J.\ High\ Energy\ Phys.} {\bf 0711}, 072 (2007)
  {\tt [arXiv:0705.4117 [hep-ph]]}.

\bibitem{delAguila:2008cj}
  F.~del Aguila and J.~A.~Aguilar-Saavedra,
  {\it Nucl.\ Phys.} {\bf B813}, 22 (2009) 
  {\tt [arXiv:0808.2468 [hep-ph]]}.

\bibitem{delAguila:2008hw}
  F.~del Aguila and J.~A.~Aguilar-Saavedra,
 {\it Phys.\ Lett.} {\bf B672}, 158 (2009) 
 {\tt [arXiv:0809.2096 [hep-ph]]}.

\bibitem{AguilarSaavedra:2009ik}
  J.~A.~Aguilar-Saavedra,
  {\tt arXiv:0905.2221 [hep-ph]}.


\bibitem{partonI}
  A.~Datta, M.~Guchait and A.~Pilaftsis,
  {\it Phys.\ Rev.} {\bf D50}, 3195 (1994) 
  {\tt [hep-ph/9311257]};
%
  F.~M.~L.~Almeida, Y.~A.~Coutinho, J.~A.~Martins Simoes and M.~A.~B.~do Vale,
  {\it Phys.\ Rev.} {\bf D62}, 075004 (2000) 
  {\tt [hep-ph/0002024]};
%
  O.~Panella, M.~Cannoni, C.~Carimalo and Y.~N.~Srivastava,
  {\it Phys.\ Rev.} {\bf D65}, 035005 (2002)
  {\tt [hep-ph/0107308]};
%
  T.~Han and B.~Zhang,
  {\it Phys.\ Rev.\ Lett.} {\bf 97}, 171804 (2006)
  {\tt [hep-ph/0604064]};
%
  S.~Bray, J.~S.~Lee and A.~Pilaftsis,
  {\it Nucl.\ Phys.} {\bf B786}, 95 (2007) 
  {\tt [hep-ph/0702294]}.


\bibitem{partonIZ}
  K.~Huitu, S.~Khalil, H.~Okada and S.~K.~Rai,
  {\it Phys.\ Rev.\ Lett.} {\bf 101}, 181802 (2008) 
  {\tt [arXiv:0803.2799 [hep-ph]]};
%
  P.~Fileviez Perez, T.~Han and T.~Li,
  {\tt arXiv:0907.4186 [hep-ph]}.


\bibitem{partonII}
  K.~Huitu, J.~Maalampi, A.~Pietila and M.~Raidal,
  {\it Nucl.\ Phys.} {\bf B487}, 27 (1997)
  {\tt [hep-ph/9606311]};
%
  J.~F.~Gunion, C.~Loomis and K.~T.~Pitts,
  {\tt [hep-ph/9610237]};
%
  A.~G.~Akeroyd and M.~Aoki,
  {\it  Phys.\ Rev.} {\bf D72}, 035011 (2005) 
  {\tt [hep-ph/0506176]};
%
  A.~Hektor, M.~Kadastik, M.~Muntel, M.~Raidal and L.~Rebane,
  {\it Nucl.\ Phys.} {\bf B787}, 198 (2007) 
  {\tt [arXiv:0705.1495 [hep-ph]]};
%
  P.~Fileviez Perez, T.~Han, G.~Y.~Huang, T.~Li and K.~Wang,
  {\it Phys.\ Rev.} {\bf D78}, 071301 (2008)
  {\tt [arXiv:0803.3450 [hep-ph]]};
%
  P.~Fileviez Perez, T.~Han, G.~y.~Huang, T.~Li and K.~Wang,
  {\it Phys.\ Rev.} {\bf D78} 015018 (2008)
  {\tt [arXiv:0805.3536 [hep-ph]]}.


\bibitem{partonIII}
  R.~Franceschini, T.~Hambye and A.~Strumia,
  {\it Phys.\ Rev.} {\bf D78}, 033002 (2008) 
  {\tt [arXiv:0805.1613 [hep-ph]]};
%
  A.~Arhrib, B.~Bajc, D.~K.~Ghosh, T.~Han, G.~Y.~Huang, I.~Puljak and G.~Senjanovic,
  {\tt arXiv:0904.2390 [hep-ph]};
%
  T.~Li and X.~G.~He,
  {\tt arXiv:0907.4193 [hep-ph]}. 


\bibitem{Sullivan:2008ki}
  Z.~Sullivan and E.~L.~Berger,
  {\it Phys.\ Rev.} {\bf D78}, 034030 (2008) 
  {\tt [arXiv:0805.3720 [hep-ph]]}.


\bibitem{LR}
  J.~C.~Pati and A.~Salam,
  {\it Phys.\ Rev.} {\bf D10}, 275 (1974) 
  [{\it Erratum-ibid.} {\bf D11}, 703 (1975)];
  R.~N.~Mohapatra and J.~C.~Pati,
  {\it Phys.\ Rev.} {\bf D11}, 2558 (1975);
  G.~Senjanovic and R.~N.~Mohapatra,
  {\it Phys.\ Rev.} {\bf D12}, 1502 (1975);
see for a review and further references 
  P.~Duka, J.~Gluza and M.~Zralek,
  {\it Annals Phys.} {\bf 280}, 336 (2000)
  {\tt [hep-ph/9910279]}.


\bibitem{Inverse}
  D.~Wyler and L.~Wolfenstein,
  {\it Nucl.\ Phys.} {\bf B218}, 205 (1983); 
  R.~N.~Mohapatra and J.~W.~F.~Valle,
  {\it Phys.\ Rev.} {\bf D34}, 1642 (1986); 
see for a definite model 
  F.~del Aguila, M.~Masip and J.~L.~Padilla,
  {\it Phys.\ Lett.} {\bf B627}, 131 (2005) 
  {\tt [hep-ph/0506063]};
and for a review and further references 
  E.~Ma,
  {\tt arXiv:0908.1770 [hep-ph]}.
 

\bibitem{recons}
  E.~J.~Chun, K.~Y.~Lee and S.~C.~Park,
  {\it  Phys.\ Lett.} {\bf B566}, 142 (2003) 
  {\tt [hep-ph/0304069]};
%
  J.~Garayoa and T.~Schwetz,
  {\it J.\ High\ Energy\ Phys.} {\bf 0803}, 009 (2008) 
  {\tt [arXiv:0712.1453 [hep-ph]]};
%
  M.~Kadastik, M.~Raidal and L.~Rebane,
  {\it Phys.\ Rev.} {\bf D77}, 115023 (2008) 
  {\tt [arXiv:0712.3912 [hep-ph]]};
%
  A.~G.~Akeroyd, M.~Aoki and H.~Sugiyama,
  {\it Phys.\ Rev.} {\bf D77}, 075010 (2008)
  {\tt [arXiv:0712.4019 [hep-ph]]};
%
  A.~G.~Akeroyd and C.~W.~Chiang,
  {\tt arXiv:0909.4419 [hep-ph]}.


\bibitem{delAguila:2008pw}
  F.~del Aguila, J.~de Blas and M.~Perez-Victoria,
  {\it Phys.\ Rev.} {\bf D78}, 013010 (2008) 
  {\tt [arXiv:0803.4008 [hep-ph]]}.

  
\bibitem{Mangano:2002ea}
  M.~L.~Mangano, M.~Moretti, F.~Piccinini, R.~Pittau and A.~D.~Polosa,
  {\it J.\ High\ Energy\ Phys.} {\bf 0307} (2003) 001
  {\tt [hep-ph/0206293].}
  
\bibitem{Sjostrand:2006za}
  T.~Sjostrand, S.~Mrenna and P.~Skands,
  {\it J.\ High\ Energy\ Phys.} {\bf 0605} (2006) 026
  [{\tt hep-ph/0603175}].
  
\bibitem{RichterWas:2002ch}
  E.~Richter-Was,
  {\tt hep-ph/0207355}.

\end{thebibliography}
\end{document}